\documentclass[11pt,a4paper]{article}

\usepackage{graphicx,xcolor}
\usepackage{amsmath}
\usepackage[numbers,sort&compress]{natbib}
\allowdisplaybreaks[2]
\definecolor{PRLblue}{RGB}{0 25 135}
\usepackage[
	bookmarks=true,
	bookmarksnumbered=true,
	colorlinks=true,
	breaklinks=true,
	citecolor=PRLblue,
	linkcolor=PRLblue,
	anchorcolor=PRLblue,
	urlcolor=PRLblue
	]{hyperref}
\usepackage{setspace}

\numberwithin{equation}{section}

\usepackage[utf8]{inputenc}
\usepackage[T1]{fontenc}
\usepackage[subscriptcorrection]{newtxmath}
\usepackage[theoremfont]{newtxtext}
\usepackage{microtype}

\usepackage{geometry}
\hypersetup{%
	pdfauthor={Tsuyoshi Houri, Makoto Sakamoto and Kentaro Tatsumi},%
	pdftitle={Spectral intertwining relations in exactly solvable quantum mechanical systems}%
}

\begin{document}

\begin{titlepage}

\begin{flushright}
{\small KOBE-COSMO-16-14, KOBE-TH-16-08\\
\today}
\end{flushright}

\vspace{1cm}

\begin{center}
\textbf{\LARGE 
\begin{spacing}{1}
Spectral intertwining relations in exactly solvable quantum-mechanical systems
\end{spacing}
}
\end{center}

\vspace{.5cm}

\begin{center}
{\large 
Tsuyoshi Houri\footnote{Email address: \texttt{houri@phys.sci.kobe-u.ac.jp}}, 
Makoto Sakamoto\footnote{Email address: \texttt{dragon@kobe-u.ac.jp}} and 
Kentaro Tatsumi\footnote{Email address: \texttt{kentaro@stu.kobe-u.ac.jp}}
}
\end{center}

\begin{center}
\textit{Department of Physics, Kobe University, Kobe 657-8501, Japan}
\end{center}

\vspace{.5cm}

\begin{abstract}%
In exactly solvable quantum-mechanical systems, ladder and intertwining operators play a central role because, if they are found, the energy spectra can be obtained algebraically. In this paper, we propose the \textit{spectral intertwining relation} as a unified relation of ladder and intertwining operators in a way that can depend on the energy eigenvalues. It is shown that the spectral intertwining relations can connect eigenfunctions of different energy eigenvalues belonging to two different Hamiltonians, which cannot be obtained by previously known structures such as shape invariance. As an application, we find new spectral intertwining operators for the Hamiltonians of the hydrogen atom and  the Rosen--Morse potential. 
\end{abstract}%

\setcounter{footnote}{0}

\end{titlepage}

\tableofcontents

\section{Introduction}\label{sec:introduction}
\begin{figure}[htbp]\centering%
	\begin{minipage}{.45\textwidth}\centering%
		\includegraphics[page=1,scale=0.62]{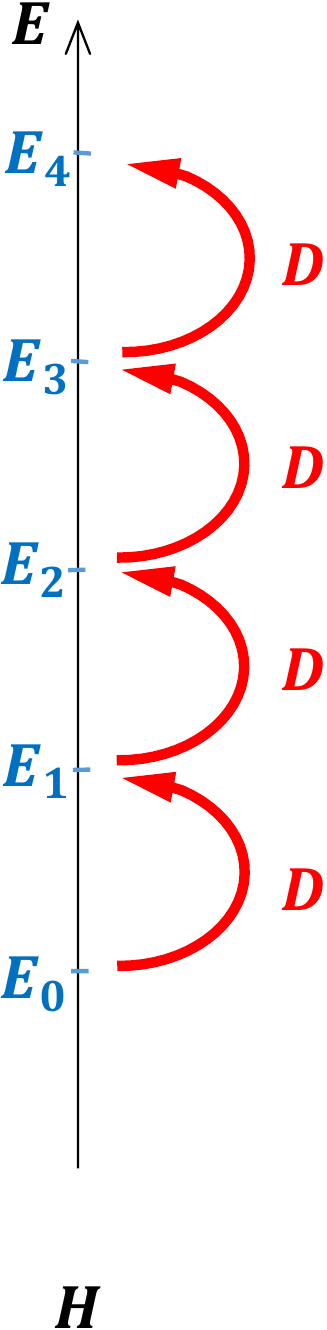}%
		\caption{The ladder structure corresponding to the commutation relation (\ref{commutation_rel_0}). The energy spectrum is equally spaced.}%
		\label{fig:fig1}%
	\end{minipage}\hspace{.1\textwidth}%
	\begin{minipage}{.45\textwidth}\centering%
		\includegraphics[page=2,scale=0.62]{figure.pdf}%
		\caption{The ladder structure corresponding to the commutation relation (\ref{str3}). The energy spectra can be unequally spaced in general.}%
		\label{fig:fig2}%
	\end{minipage}%
\end{figure}

%
What is an important structure commonly found in exactly solvable quantum-mechanical systems? The concept of ladder operators is of great interest in this question because, since Dirac's use \cite{Dirac:1930}, it has played a central role in quantum mechanics. As is well known, a ladder operator $D$ for a Hamiltonian $H$ is a linear operator that satisfies the commutation relation
\begin{align}
	[H,D] & = \varepsilon D \,,
\label{commutation_rel_0}
\end{align}
where $\varepsilon$ is a real constant. In particular, it is called a raising operator if $\varepsilon>0$ or a lowering operator if $\varepsilon<0$ as it raises or lowers the energy eigenvalues. In fact, when $D$ acts on an eigenfunction $\psi_n$ with energy $E_n$, i.e., $H \psi_n = E_n \psi_n$, we see that
\begin{align}
	H D \psi_n & = D H \psi_n + [H,D] \psi_n  \nonumber\\
	& = (E_n + \varepsilon ) D \psi_n \,,
\end{align}
which shows that $D\psi_n$ is an eigenfunction with energy $E_n+\varepsilon$. (Note: In this paper, we focus only on bound states, so that the energy eigenfunctions $\psi_n$ and energy eigenvalues $E_n$ are discretized and labeled by integers $n$. We assign $0$ to the ground state and $1,2,\dots$ to the excited states, unless otherwise noted.) Hence, for any $n$, there exists an integer $m$ such that $\psi_{n+m} \propto D\psi_n$ and $E_{n+m} = E_n + \varepsilon$. Since it follows that, if $m=1$, $\psi_n\propto D^n\psi_0$ and $E_n = E_0 + n\varepsilon$, ladder operators allow us to construct the full energy spectrum for the bound states algebraically (Fig.\ \ref{fig:fig1}). However, we notice that the commutation relation (\ref{commutation_rel_0}) always leads to equally spaced energy spectra, while almost all the systems that we know of have energy spectra that are not equally spaced. This motivates us to extend the commutation relation (\ref{commutation_rel_0}) in a way to enable us to construct unequally spaced energy spectra algebraically. A remarkable formulation for such ladder operators has been proposed as the closure relation \cite{Odake:2006pl,Odake:2006jmp,Odake:2016}, which is the double commutation relation for a Hamiltonian and a function called a sinusoidal coordinate. Given a sinusoidal coordinate creation/annihilation operators for energy eigenstates can be constructed from it, and the energy spectra generated by them become unequally spaced in general, as shown in Fig.\ \ref{fig:fig2}.

\begin{figure}[htbp]\centering%
	\begin{minipage}{.85\textwidth}\centering%
		\includegraphics[page=3,scale=0.62]{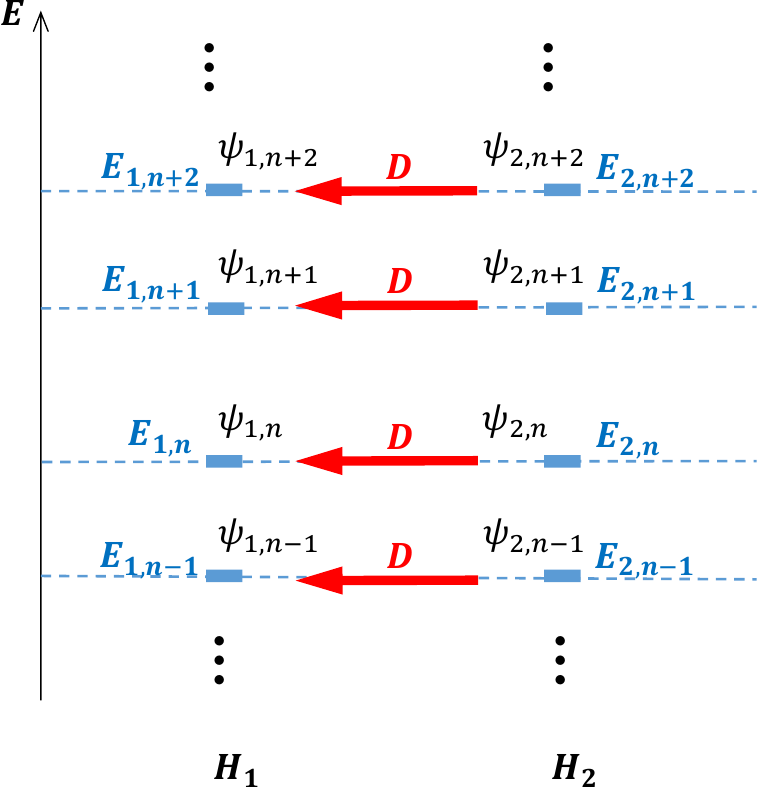}%
		\caption{A sketch of the intertwining relation (\ref{intertwining_rel_0}). The operator $D$ connects eigenfunctions for $H_2$ into eigenfunctions for $H_1$ in the same energy level.}%
		\label{fig:fig3}%
	\end{minipage}%
\end{figure}

In this paper we also study intertwining operators in the context of quantum mechanics; they have already been studied in many areas of mathematics and physics. An intertwining operator $D$ in quantum mechanics is a linear operator that satisfies the intertwining relation
\begin{align}
	H_1 D & = D H_2 \,, \label{intertwining_rel_0}
\end{align}
where $H_1$ and $H_2$ are two different Hamiltonians.
From this relation, it can be found that for an eigenfunction $\psi_{2,n}$ for $H_2$ obeying $H_2\psi_{2,n} = E_{2,n} \psi_{2,n}$,
\begin{align}
	H_1 \left(D\psi_{2,n}\right) &
	= D \left(H_2\psi_{2,n}\right) 
	= E_{2,n} \left(D\psi_{2,n}\right) \,,
\end{align}
which shows that $D\psi_{2,n}$ is an eigenfunction for $H_1$. Thus the intertwining operator $D$ maps eigenfunctions for $H_2$ into eigenfunctions for $H_1$, as shown in Fig.\ \ref{fig:fig3}. 

\begin{figure}[htbp]\centering%
	\begin{minipage}{.85\textwidth}\centering%
		\includegraphics[page=4,scale=0.62]{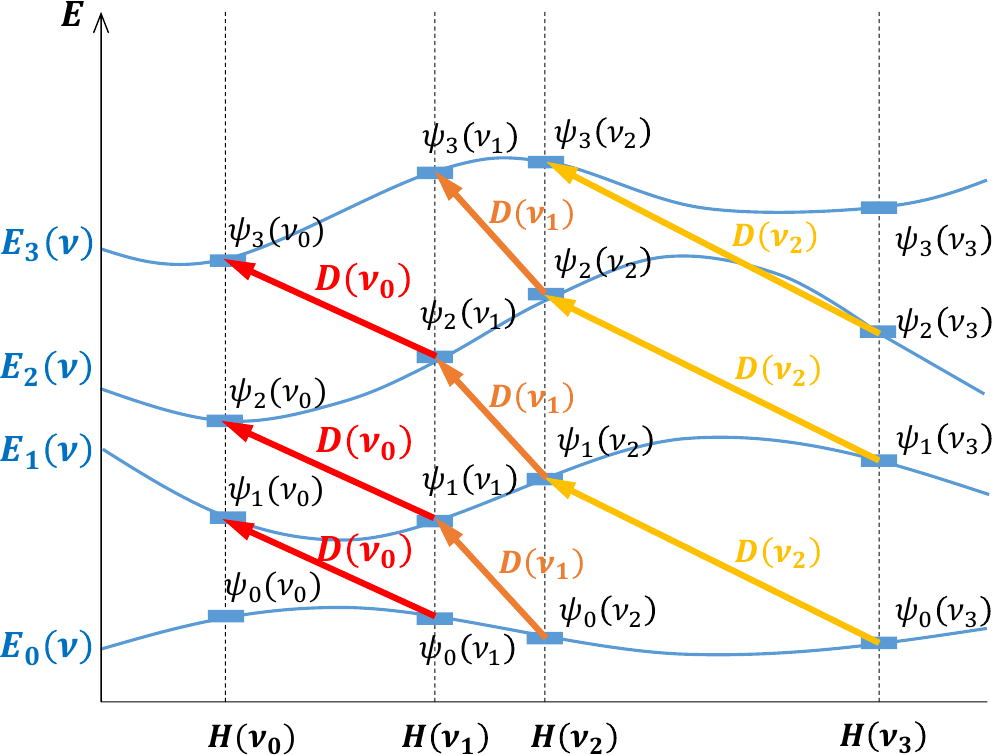}%
		\caption{A diagram describing how the spectral intertwining operators $D(\nu)$ satisfying Eq.\ (\ref{basic_eq}) map energy eigenfunctions for $H(\nu_i)$, where $\nu_i= f(f(\cdots f(f(\nu_0))\cdots))$ for $i=1,2,\dots$. Since the shifts of energy eigenvalues depend only on $\nu$, all the arrows between $H(\nu_{i-1})$ and $H(\nu_i)$ are parallel.}%
		\label{fig:label4}%
	\end{minipage}%
\end{figure}

\begin{figure}[htbp]\centering%
	\begin{minipage}{.85\textwidth}\centering%
		\includegraphics[page=5,scale=0.62]{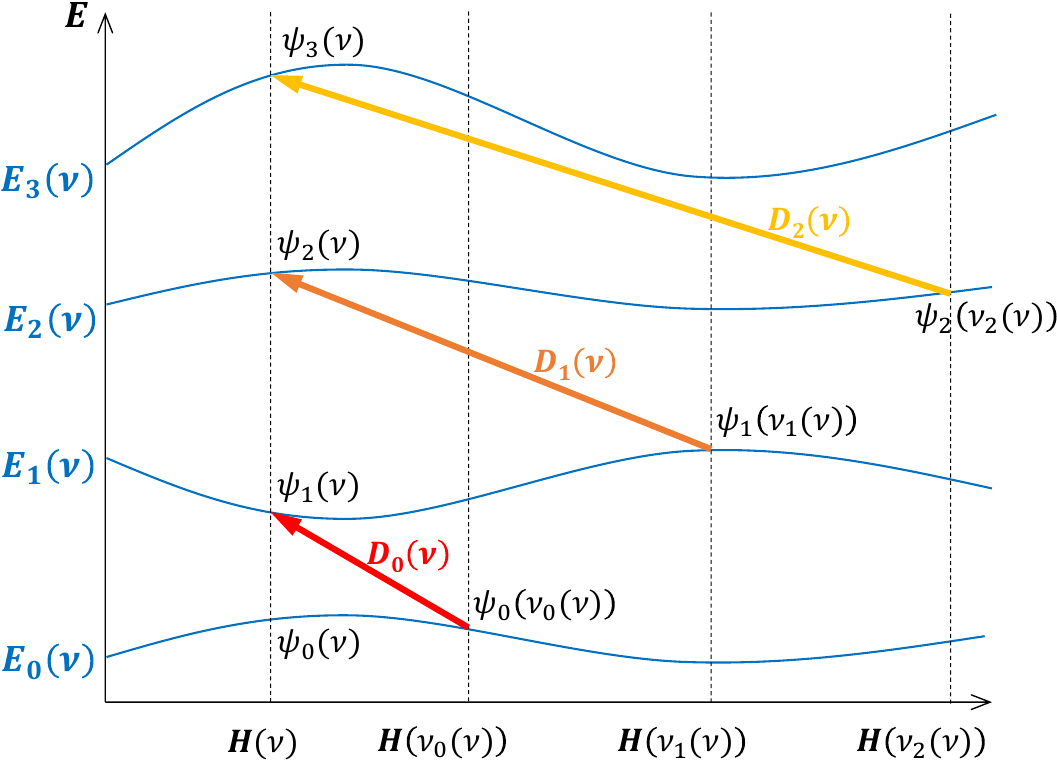}%
		\caption{A diagram describing how the spectral intertwining operators $D_n(\nu)$ act on the energy eigenfunctions $\psi_n\left(\nu_n(\nu)\right)$ for $H\left(\nu_n(\nu)\right)$ for $n=0,1,2,\dots$.}%
		\label{fig:label5}%
	\end{minipage}%
\end{figure}

The two concepts of ladder and intertwining operators are basically different. However, when a Hamiltonian is parametrized by a parameter or a set of parameters, we can organize the two concepts in a unified way as \textit{spectral intertwining operators}. The linear operators connect eigenfunctions of different eigenvalues between two different Hamiltonians, as shown in Figs.\ \ref{fig:label4} and \ref{fig:label5}. In this paper, we point out that the presence of spectral intertwining operators is important for quantum-mechanical systems being exactly solvable. In supersymmetric quantum mechanics \cite{Witten:1981}, spectral intertwining operators have already been studied as shape invariance \cite{Gendenshtein:1983}. In supersymmetric quantum mechanics , two supersymmetric Hamiltonians are related to each other by the intertwining relations
\begin{align}
	H_+ a^\dag &= a^\dag H_- \,,& H_-a &= a H_+ \,, 
\label{intertwining_relation}
\end{align}
where $a^\dag, a$ are intertwining operators (called supercharges since $a^\dag$ ($a$) maps eigenfunctions for $H_-$ ($H_+$) into eigenfunctions for $H_+$ ($H_-$) at the same energy level). In addition, when two Hamiltonians have shape invariance, the energy spectrum of the system can be solved exactly by using the supercharges in combination with shape invariance (see Sect.\ \ref{sec:sip} for details). In the past, many exactly solvable quantum-mechanical systems have been studied by means of shape invariance (see, e.g., Ref.\ \cite{Cooper:1995}). While the shape invariance is very powerful to construct exactly solvable quantum-mechanical models, it is important to obtain other concepts to characterize such models. Very recently, a coupled system in supersymmetric quantum mechanics was studied in  Ref.\ \cite{Williams:2017}, where the authors discussed the underlying structure for operators that connect eigenstates between two sectors.

In previous works, the operators explained above have been formulated in $n$-independent ways. They have a merit that one does not have to know the $n$-dependence of the eigenfunctions and eigenvalues. In contrast, we dare to formulate them in an $n$-dependent way in the present paper because this could allow us to study as many examples as possible or to find new examples that have never been elucidated. We attempt to make a formulation for spectral intertwining operators. Actually, we show several examples in Sect.\ \ref{sec:examples_SIR}. Previously known formulations such as the closure relation and shape invariance are discussed in the appendices, which will illuminate how special they are in our $n$-dependent formulation.

This paper is organized as follows: In the next section, we propose the spectral intertwining relations by unifying the relation of ladder and intertwining relations. In Sect.\ \ref{sec:examples_SIR}, as an example, we show spectral intertwining relations for several Hamiltonians such as the hydrogen atom (Coulomb potential) and the Rosen--Morse potentials. It is known that the Coulomb and Rosen--Morse potentials admit spectral intertwining operators associated with the shape invariance. However, the spectral intertwining operators that we obtain are different from them. Section \ref{sec:SandD} is devoted to a summary and discussion. In Appendix \ref{sec:ladder_op}, we consider the one-parameter representation for ladder operators, which leads to a modification of the commutation relation (\ref{commutation_rel_0}) in a natural way. Several examples that fit into the modified commutation relation are also given. In Appendix \ref{sec:sec_3}, we discuss the case in which a Hamiltonian is parametrized and a spectral intertwining relation is introduced. It is also shown that the relation introduced can be found in supersymmetric quantum mechanics with shape invariance.

\section{Spectral intertwining relation}\label{sec:General_SIR}%
When a Hamiltonian $H(\nu)$ is parametrized by a parameter $\nu$ (or a set of parameters), the Schr\"odinger equation is given by
\begin{align}
	H(\nu) \psi_n(\nu) & = E_n(\nu) \psi_n(\nu) \,.
\label{top_of_2_1}
\end{align}
For such a Hamiltonian, we propose a \textit{spectral intertwining relation} as the $n$-dependent commutation relation
\begin{align}
	H(\nu) D_n(\nu) & = D_n(\nu) \big(H\left(\nu_n(\nu)\right)+\varepsilon_n\left(\nu_n(\nu)\right)\big) \nonumber\\
	&\quad + Q_n\left(\nu_n(\nu)\right) \big(H\left(\nu_n(\nu)\right) - E_n\left(\nu_n(\nu)\right)\big) \,,
\label{general_type}
\end{align}
where $\nu$ and $n$ are parameters, $Q_n\left(\nu_n(\nu)\right)$ are some operators, and $\varepsilon_n\left(\nu_n(\nu)\right)$, $E_n\left(\nu_n(\nu)\right)$ and $\nu_n(\nu)$ are constants depending on $\nu$ and $n$. We call $D_n(\nu)$ a \textit{spectral intertwining operator}.

Given $D_n(\nu)$ satisfying the commutation relation (\ref{general_type}) and an eigenfunction $\psi_n\left(\nu_n(\nu)\right)$ for $H\left(\nu_n(\nu)\right)$, we obtain
\begin{align}
	H(\nu) D_n(\nu) \psi_n\left(\nu_n(\nu)\right) & =  \big(E_n\left(\nu_n(\nu)\right)+\varepsilon_n\left(\nu_n(\nu)\right)\big) D_n(\nu) \psi_n\left(\nu_n(\nu)\right) \,,
\label{eq69}
\end{align}
which shows that $D_n(\nu) \psi_n\left(\nu_n(\nu)\right)$ is an eigenfunction for the Hamiltonian $H(\nu)$ with energy $E_n\left(\nu_n(\nu)\right)+\varepsilon_n\left(\nu_n(\nu)\right)$. Hence, for any $n$, there exists an integer $m$ such that
\begin{align}
	\psi_{n+m}(\nu) &\propto D_n(\nu) \psi_n\left(\nu_n(\nu)\right) \,,
\label{lad_act_2}\\
	E_{n+m}(\nu) &= E_n\left(\nu_n(\nu)\right)+\varepsilon_n\left(\nu_n(\nu)\right) \,,
\label{lad_act_2_2}
\end{align}
when $\|D_n(\nu)\psi_n\left(\nu_n(\nu)\right)\|<\infty$ and $D_n(\nu)\psi_n\left(\nu_n(\nu)\right)\neq0$. If $m=1$ for all $n$, as before, we get the diagram shown in Fig.\ \ref{fig:label5}. (For nontrivial examples of $m\neq1$, see Ref.\ \cite{Odake:2016}.) As mentioned in the previous section, these operators connect eigenfunctions of different eigenvalues between two Hamiltonians. The difference from the previous ones is that such spectral intertwining operators depend not only on $\nu$, which parametrizes the theories, but also on $n$, which parametrizes the eigenfunctions and eigenvalues. In the next section, we provide some examples of Hamiltonians that admit such spectral intertwining operators.

\section{Examples of the spectral intertwining relation}\label{sec:examples_SIR}
In this section, we give examples of the spectral intertwining relations of the hydrogen atom and the Rosen--Morse potentials. Although these models have shape invariance, the spectral intertwining relations cannot be derived from it.

\subsection{Hydrogen atom}\label{sec:examples_HA}

The Hamiltonian is parametrized by the mass $m$ and the coupling constants $\varg$ and $\ell$ as
\begin{align}
	H(m,\varg,\ell)&\equiv \frac{1}{2m}\left[-\frac{1}{r^2}\frac{d}{dr}\left(r^2\frac{d}{dr}\right) + \frac{\ell(\ell+1)}{r^2}-\frac{2\varg}{r}\right] \,,&
	0&\leq r<\infty \,,
\label{Hamiltonian_2_2}
\end{align}
which recovers the radial part of the Schr\"{o}dinger equation for the hydrogen atom if $m=m_e$, $\varg=m_e\varg_0$, and $\ell=0,1,2,\dots$.  Since the Hamiltonian depends on $m$, $\varg$, and $\ell$, the eigenfunctions and eigenvalues also depend on them in general. In the present case, we have
\begin{align}
	\psi_n(\varg,\ell) & =C_{\varg,n,\ell} r^\ell \exp\left(-\frac{\varg r}{n+\ell+1}\right)
		L_{n}^{2\ell+1}\left(\frac{2\varg r}{n+\ell+1}\right) \,, \\
	E_n(m,\varg,\ell) & = - \frac{\varg^2}{2m(n+\ell+1)^2} \,,
	\qquad n=0,1,\dots,
\end{align}
where $C_{\varg,n,\ell}$ are normalization constants and $L_n^\alpha(\xi)$ are the associated Laguerre functions in $\xi$.

In a heuristic way, we are able to find the operator
\begin{align}
	D_n(\varg,\ell) & \equiv r\frac{d}{dr} - \frac{\varg r}{n+\ell+2} + n+\ell+2 \,,
\label{sio_HO_2}
\end{align}
which satisfies the spectral intertwining relation
\begin{align}
	H(m,\varg,\ell)D_n(\varg,\ell) & = D_n(\varg,\ell)H\left(m,\varg_n(\varg,\ell),\ell\right)+ 2\left(H\left(m,\varg_n(\varg,\ell),\ell\right) + \frac{\varg_n(\varg,\ell)^2}{2m(n+\ell+1)^2} \right) \,,
\label{sio_HO_2_relation}
\end{align}
where
\begin{align}
	\varg_n(\varg,\ell) & = \varg \frac{n+\ell+1}{n+\ell+2} \,.
\end{align}
This relation fits into the spectral intertwining relation (\ref{general_type}) with
\begin{align}
	\varepsilon_n(m,\varg_n(\varg,\ell),\ell)&=0\,,&
	Q_n(m,\varg_n(\varg,\ell),\ell)&=2\,,&
	E_n(m,\varg_n(\varg,\ell),\ell)&=-\frac{\varg_n(\varg,\ell)^2}{2m(n+\ell+1)^2}\,.
\end{align}
$D_n(\varg,\ell)$ maps an eigenfunction $\psi_n\left(\varg_n(\varg,\ell),\ell\right)$ into an eigenfunction $\psi_{n+1}(\varg,\ell)$. Since we are now interested in the Hamiltonian $H(m,\varg,\ell)$
, we obtain
\begin{align}
	\psi_{n+1}(\varg,\ell) & \propto D_n(\varg,\ell) \psi_n\left(\varg_n(\varg,\ell),\ell\right) \,.
\label{Shift_rel}
\end{align}


A further insight is obtained by rescaling the coordinate $r$ in the Hamiltonian \eqref{Hamiltonian_2_2}. Rescaling $r$ with a parameter $\alpha$, we easily find that
\begin{align}
	H(m,\varg,\ell;\alpha r) &	= \frac{1}{\alpha^2}H(m,\alpha \varg,\ell; r)
		= H(\alpha^2 m,\alpha \varg,\ell; r) \,.
\end{align}
This implies that the energy eigenfunctions of the Hamiltonian \eqref{Hamiltonian_2_2} rescaled by $r\to \alpha r$ are equivalent to those rescaled by $m\to\alpha^2 m$ and $\varg\to \alpha \varg$ up to normalization, i.e., 
\begin{align}
	\psi_n(\varg,\ell;\alpha r) & \propto \psi_n(\alpha \varg,\ell; r) \,.
\label{Shift_rel_2}
\end{align}
Since the rescaling of the coordinate $r$ can be represented by the operator $S(\alpha) = \exp\left[(\ln \alpha) r d/dr\right]$, we have
\begin{align}
	S(\alpha)\psi_n(\varg,\ell;r) & = \psi_n(\varg,\ell;\alpha r) \,.
\label{Shift_rel_3}
\end{align}
Combining (\ref{Shift_rel}), (\ref{Shift_rel_2}) and (\ref{Shift_rel_3}) with $\alpha_n(\ell) \varg\equiv \varg_n(\varg,\ell) = (n+\ell+1)/(n+\ell+2)\varg$, we obtain
\begin{align}
	\psi_{n+1}(\varg,\ell) & \propto D_n\left(\varg,\ell\right)S\left(\alpha_n(\ell)\right)\psi_n(\varg,\ell) \,.
\end{align}
Since we have
\begin{align}
	H(m,\varg,\ell)D_n(\varg,\ell) & = D_n(\varg,\ell)H\left(m,\alpha_n(\ell) \varg,\ell\right)\nonumber\\
		&\quad+ 2\left(H\left(m,\alpha_n(\ell) \varg,\ell\right) - E_{n}\big(m/\alpha_n(\ell)^2,\varg\big) \right) \,,\\
	H\left(m,\alpha_n(\ell)\varg,\ell\right)S\left(\alpha_n(\ell)\right)
		&= \alpha_n(\ell)^2 S\left(\alpha_n(\ell)\right)H(m,\varg,\ell) = S\left(\alpha_n(\ell)\right) H\big(m/\alpha_n(\ell)^2,\varg,\ell\big)\,,
\end{align}
the composite operator
\begin{align}
	\tilde{D}_n(\varg,\ell) &\equiv D_n(\varg,\ell)S\left(\alpha_n(\ell)\right)
\label{sio_HO_3}
\end{align}
satisfies the spectral intertwining relation
\begin{align}
	H(m,\varg,\ell)\tilde{D}_n(\varg,\ell) & = \tilde{D}_n(\varg,\ell)H\big(m/\alpha_n(\ell)^2,\varg,\ell\big)\nonumber\\
		&\quad+2S\left(\alpha_n(\ell)\right)\left(H\big(m/\alpha_n(\ell)^2,\varg,\ell\big) - E_{n}\big(m/\alpha_n(\ell)^2,\varg\big)\right) \,.
\label{sio_HO_3_relation}
\end{align}

We have found that the system of the hydrogen atom possesses the spectral intertwining relations (\ref{sio_HO_2_relation}) and (\ref{sio_HO_3_relation}), which are related to the parameters $\ell$, $\varg$, and $m$. These operators had already been found in Refs. \ \cite{Musto:1966,Pratt:1966,Malkin:1966,Barut:1967a,Barut:1967b} as part of the dynamical group $O(4,2)$, but they have not been viewed as spectral intertwining operators before.

\subsection{Rosen--Morse potential}\label{sec:examples_RM}
%

The 1D Hamiltonian with the spherical (or trigonometric) Rosen--Morse potential is given by
\begin{align}
	H_\text{sph.}(\varg,\ell) & =-\frac{d^{2}}{dx^{2}}+\frac{\ell(\ell+1)}{\sin^{2}x}-2\varg\cot x\,, & 
	0& \leq x\leq \pi \,,
\end{align}
where $0\leq \ell$ and $0\leq \varg$ are parameters. The eigenfunctions and eigenvalues are given by
\begin{align}
	\psi_n(\varg,\ell;x)& =C_{\varg,\ell,n}
		(\sin x)^{n+l+1}e^{-\frac{\varg}{n+\ell+1}x}
		P_{n}^{\left(a_{+}(\varg,\ell,n),a_{-}(\varg,\ell,n)\right)}(i\cot x)\,,\\
	E_n(\varg,\ell) & =+(n+\ell+1)^{2}-\left(\frac{\varg}{n+\ell+1}\right)^{2}\,,
\end{align}
where $C_{\varg,\ell,n}$ are the normalization constants, $P_n^{(\alpha,\beta)}(\xi)$ are the Jacobi functions in $\xi$ and $a_{\pm}(\varg,\ell,n)$ are
\begin{align}
	a_{\pm}(\varg,\ell,n) & =-(n+\ell+1)\pm i\frac{\varg}{n+\ell+1}\,.
\end{align}
The spectral intertwining operators are given by
\begin{align}
	D_n(\varg,\ell)& =\frac{\tilde{a}_{+}-\tilde{a}_{-}}{2i} \sin x+ \frac{\tilde{a}_{+}+\tilde{a}_{-}}{2}\cos x - \sin x \frac{d}{dx} \,,
\end{align}
where
\begin{align}
	\tilde{a}_{\pm}(\varg,\ell,n)& =a_{\pm}\left(\varg_n(\varg,\ell),\ell,n\right)=a_{\pm}\left(\varg\frac{n+\ell+1}{n+\ell+2},\ell,n\right)\,,&
	\varg_n(\varg,\ell)&\equiv \varg\frac{n+\ell+1}{n+\ell+2}\,.
\end{align}
They satisfy the spectral intertwining relation
\begin{align}
	H_\text{sph.}(\varg,\ell)D_n(\varg,\ell)&=D_n(\varg,\ell)\Big(H_\text{sph.}\left(\varg_n(\varg,\ell),\ell\right)+2(n+\ell+1)+1\Big) \nonumber\\
	&\quad-2\cos x \Big(H_\text{sph.}\left(\varg_n(\varg,\ell),\ell\right)-E_n\left(\varg_n(\varg,\ell),\ell\right)\Big) \,,
\label{sph_RM_SIR}
\end{align}
which fits into Eq.\ (\ref{general_type}) by setting
\begin{align}
	\varepsilon_n\left(\varg_n(\varg,\ell),\ell\right)&\equiv2(n+\ell+1)+1\,,&
	Q_n\left(\varg_n(\varg,\ell),\ell\right)&\equiv-2\cos x	\,.
\end{align}


Next, we consider the 1D Hamiltonian with the hyperbolic Rosen--Morse potential
\begin{align}
	H_\text{hyp.}(\varg,\ell) & =-\frac{d^{2}}{dx^{2}}+\frac{\ell(\ell+1)}{\sinh^{2}x}-2\varg\coth x \,,
	&0& \leq x<\infty \,,\label{eq:hamiltonian_hyp_RM}
\end{align}
where $0\leq \ell$ and $(\ell+1)^2\leq \varg$ are parameters. The eigenfunctions and eigenvalues are given by
\begin{align}
	\psi_n(\varg,\ell;x)& =C_{\varg,\ell,n}(\sinh x )^{n+\ell+1}e^{-\frac{\varg}{n+\ell+1}x}
	P_{n}^{\left(b_{+}(\varg,\ell,n),b_{-}(\varg,\ell,n)\right)}(\coth x)\,,\\
	E_n(\varg,\ell) & =-(n+\ell+1)^{2}-\left(\frac{\varg}{n+\ell+1}\right)^{2}\,,
\end{align}
where $C_{\varg,\ell,n}$ are the normalization constants and $P_n^{(\alpha,\beta)}(\xi)$ are the Jacobi functions in $\xi$. We note that the system (\ref{eq:hamiltonian_hyp_RM}) has finitely many discrete eigenstates $\psi_n(\varg,\ell;x)$ with $0 \leq n < \sqrt{\varg}-\ell-1$. For the hyperbolic potential, $b_{\pm}$ are given by
\begin{align}
	b_{\pm}(\varg,\ell,n) & =-(n+\ell+1)\pm\frac{\varg}{n+\ell+1}\,.
\end{align}
The spectral intertwining operators are given by
\begin{align}
	D_n(\varg,\ell)& =\frac{\tilde{b}_{+}-\tilde{b}_{-}}{2}\sinh x + \frac{\tilde{b}_{+}+\tilde{b}_{-}}{2}\cosh x - \sinh x\frac{d}{dx}  \,,
\end{align}
where
\begin{align}
	\tilde{b}_{\pm}(\varg,\ell,n)&=b_{\pm}(\varg_n(\varg,\ell),\ell,n)=b_{\pm}\left(\varg\frac{n+\ell+1}{n+\ell+2},\ell,n\right)\,,&
	\varg_n(\varg,\ell)&\equiv \varg\frac{n+\ell+1}{n+\ell+2}\,.
\end{align}
They satisfy the spectral intertwining relation
\begin{align}
	H_\text{hyp.}(\varg,\ell)D_n(\varg,\ell)&=D_n(\varg,\ell)\Big(H_\text{hyp.}\left(\varg_n(\varg,\ell),\ell\right)-2(n+\ell+1)-1\Big) \nonumber\\
	&\quad-2\cosh x \Big(H_\text{hyp.}\left(\varg_n(g,\ell),\ell\right)-E_n\left(\varg_n(\varg,\ell),\ell\right)\Big) \,,\label{hyp_RM_SIR}
\end{align}
which fits into Eq.\ (\ref{general_type}) by setting
\begin{align}
	\varepsilon_n\left(\varg_n(\varg,\ell),\ell\right)&\equiv-2(n+\ell+1)-1\,,&
	Q_n\left(\varg_n(\varg,\ell),\ell\right)&\equiv -2\cosh x	\,.
\end{align}

It is known that these Rosen--Morse potentials have shape invariance which leads to the spectral intertwining relation with respect to $\ell$.
In contrast, here, we have obtained the spectral intertwining relation with respect to $\varg$ and $\ell$, which cannot be derived from shape invariance.

\section{Summary and discussion}\label{sec:SandD}

In this paper we have considered the question of what is an important structure commonly found in exactly solvable quantum-mechanical systems. For a parametrized Hamiltonian, we have proposed the spectral intertwining relation (\ref{general_type}) as a unified relation of ladder and intertwining relations.
Since it is formulated in an $n$-dependent way, the spectral intertwining relations can connect eigenfunctions between two different Hamiltonians.
It is emphasized that each of the spectral intertwining relations (\ref{sio_HO_2_relation}), (\ref{sio_HO_3_relation}), (\ref{sph_RM_SIR}) and (\ref{hyp_RM_SIR}) connects two different Hamiltonians, which cannot be obtained by shape invariance. Thus, our formulation can deal with many examples including previously known ones, and can connect many models in a wider frame.

It has been thought that ladder and intertwining operators premise the (dynamical) symmetry of a system, so that they have been studied in algebraic ways. For example, shape invariance in supersymmetric quantum mechanics can be interpreted by means of extended Lie algebras \cite{Balantekin:1998,Gangopadhyaya:1998}. Hence, as it was pointed out in Refs.\ \cite{Musto:1966,Pratt:1966,Malkin:1966,Barut:1967a,Barut:1967b} that the spectral intertwining operators for the hydrogen atom are part of the dynamical group $O(4,2)$, we expect that the spectral intertwining relation (\ref{general_type}) is in general related to some sort of symmetry of a system. To see this, it would be interesting to lift a quantum system to a spacetime with Bergmann structure in higher dimensions \cite{Duval:1985,Duval:1994a,Duval:1994b,Cariglia:2016}, where the Schr\"odinger symmetry in the original quantum system can be found explicitly as part of the conformal symmetry in the lifted spacetime. Some other lifts have also been studied in Refs. \cite{Karamatskou:2014,Cariglia:2015}. In this way, the spectral intertwining relation (\ref{general_type}) would be of interest from a geometric point of view.

\section*{Acknowledgements}
The authors would like to thank Masato Nozawa and Satoru Odake for useful comments. This work was supported in part by JSPS KAKENHI Grant No. JP14J01237 (T.H.) and Grants-in-Aid for Scientific Research (No. 15K05055 and No. 25400260 (M.S.)) from the Ministry of Education, Culture, Sports, Science and Technology (MEXT) in Japan.

\newpage%
\appendix%
%
\section{One-parameter representation of ladder operators}\label{sec:ladder_op}

In this section, we consider how to modify the commutation relation (\ref{commutation_rel_0}). In order to organize unequally spaced energy spectra, we make the ladder operators depend on the energy level $n$, which can be done with the commutation relation
\begin{align}
	[H,D(a)] & = \varepsilon(a) D(a) + Q(a) \left(H - E(a)\right) \,,
\label{str3}
\end{align}
where $a$ is a parameter, $\varepsilon(a)$ and $E(a)$ are real constants for every $a$, and $Q(a)$ is a linear operator depending on $a$.  We note that, if $D(a)$ and $\varepsilon(a)$ are independent of the parameter $a$ and $Q(a)\equiv 0$, it recovers the commutation relation (\ref{commutation_rel_0}).

Given a linear operator $D(a)$ satisfying the commutation relation (\ref{str3}) for a Hamiltonian $H$, we set $a=a_n$ so as to satisfy
\begin{align}
	E_n &= E(a_n) \,,&
	D_n&= D(a_n) \,,&
	\varepsilon_n&=\varepsilon(a_n) \,,&
	n&=0,1,2\dots
\end{align}
and obtain for an eigenfunction $\psi_n$ when $\|D_n\psi_n\|<\infty$ and $D_n\psi_n\neq0$,
\begin{align}
	H D_n\psi_n &= (E_n + \varepsilon_n) D_n\psi_n \,,&
	n&=0,1,2\dots
\label{cond_1}
\end{align}
which shows that $D_n\psi_n$ is an eigenfunction with energy $E_n + \varepsilon_n$. Hence, for any $n$, there exists an integer $m$ such that $\psi_{n+m}\propto D_n\psi_n$ and $E_{n+m} = E_n + \varepsilon_n$; i.e., $D_n$ acts as a ladder operator for an eigenfunction $\psi_n$. We should note that the value $a_n$ must be chosen appropriately so as to satisfy the condition (\ref{cond_1}). If the value $a_n$ is not appropriate, $D_n\psi_n$ does not become an eigenfunction in general. If $m=1$ for all $n$, the energy spectrum can be constructed entirely in the manner shown in Fig.\ \ref{fig:fig2}, and we obtain, for $n=1,2,\dots$,
\begin{align}
	\psi_n &\propto D_{n-1}\cdots D_0 \psi_0 \,, &
	E_n & = E_0 + \sum_{i=0}^{n-1} \varepsilon_i \,,
\end{align}
where we emphasize again that each $a_i$ satisfies the condition (\ref{cond_1}), i.e., $E(a_i)=E_i$ $(i=0,1,2,\dots)$. Thus it is sufficient to obtain the eigenfunction $\psi_0$ for the ground state to construct the full energy spectrum.

In general, we may replace the second term on the right-hand side of Eq.\ (\ref{str3}) with a more general form, e.g., replacing $f(H-E_n)$ with an arbitrary function $f(x)$ with $f(0)=0$. However, since we want to get ladder operators as differential operators, it is reasonable to assume that $f$ is polynomial, which leads to the current form in the end.

It should also be noted that, if the energy eigenstates are degenerate, additional quantum numbers should be introduced. Accordingly, spectral intertwining operators can have several parameters. For example, if there exists a linear operator $K$ that commutes with a Hamiltonian $H$, $[H,K]=0$, the commutation relation (\ref{str3}) can be modified to
\begin{align}
	[H,D(a,b)] & = \varepsilon(a,b) D(a,b) + Q_1(a,b) (H - E(a) ) + Q_2(a,b)(K - \lambda(b))\,.
\label{str3_2}
\end{align}

\subsection{Harmonic oscillator in one dimension}\label{sec:harmonic-oscillator}

The Schr\"odinger equation for a harmonic oscillator in one dimension is given by the Hamiltonian
\begin{align}
	H_\text{HO} & \equiv -\frac{1}{2m}\frac{d^2}{dx^2} + \frac{1}{2}m\omega^2x^2 \,, &
	-\infty&<x<\infty \,,
\label{Hamiltonian_HO}
\end{align}
where $m>0$ and $\omega$ are the mass and frequency, respectively. The eigenfunctions and eigenvalues are given by
\begin{align}
	\psi_n &= C_n \exp\left(-\frac{m\omega x^2}{2}\right) H_n\left(\sqrt{m\omega}x\right) \,, \\
	E_n &= \left(n+\frac{1}{2}\right)\omega \,, \qquad
	n=0,1,2,\dots,
\end{align}
where $C_n$ are normalization constants and $H_n(\xi)$ are the Hermite polynomials in $\xi$. The well known ladder operators are given by
\begin{align}
	a_\pm & \equiv \mp \sqrt{\frac{1}{2m\omega}}\left(\frac{d}{dx} \mp m\omega x\right) \,,
\label{ladder_HO_1}
\end{align}
which satisfy the commutation relations
\begin{align}
	[H_\text{HO},a_\pm] & =\pm \omega a_\pm \,.
\label{commutation_rel_HO}
\end{align}

It is well known that the commutation relations (\ref{commutation_rel_HO}) are derived from $H_\text{HO}=(a_+a_-+1/2)\omega$ and the Heisenberg algebra $[a_-,a_+]=1$. Using the Heisenberg algebra, it is also shown that the linear operators $\hat{D}_\pm \equiv \mp a_\pm a_\pm$ satisfy $[H_\text{HO},\hat{D}_\pm] = \pm 2\omega \hat{D}_\pm$, which shows that $\hat{D}_\pm$ are ladder operators that map $\psi_n$ into $\psi_{n\pm2}$. As differential operators, they are explicitly written as
\begin{align}
	\hat{D}_\pm & \equiv x \frac{d}{dx} \mp m\omega x^2 \pm \frac{H_\text{HO}}{\omega}+\frac{1}{2} \,,
\label{Dhat}
\end{align}
where $H_\text{HO}$ is the Hamiltonian (\ref{Hamiltonian_HO}). Since $H_\text{HO}$ is replaced by the energy eigenvalue $E_n$ when $\hat{D}_\pm$ act on an eigenfunction $\psi_n$, $\hat{D}_\pm$ may be expressed as first-order operators in the form
\begin{align}
	D_n^{(\pm)} & \equiv x \frac{d}{d x} \mp m\omega x^2 \pm \left(n+\frac{1}{2}\right)+\frac{1}{2} \,,
\label{ladder_HO_2}
\end{align}
where $n$ is an integer. $D_\pm(n)$ satisfy the commutation relations
\begin{align}
	[H_\text{HO},D_n^{(\pm)}] & = \pm 2 \omega D_n^{(\pm)} + 2\left(H_\text{HO}-E_n \right) \,,
\label{commutation_rel_HO2}
\end{align}
which fit into the commutation relation (\ref{str3}).

In the process from (\ref{Dhat}) to (\ref{ladder_HO_2}) the second-order operators $\hat{D}_\pm$ were transformed into the first-order operators $D_n^{(\pm)}$ by replacing the Hamiltonian $H_\text{HO}$ with the eigenvalues $E_n$. In return for this, $D_n^{(\pm)}$ came to depend on $n$. They are, of course, just different representations. However, when we want to get expressions of ladder operators as differential operators, the latter is more useful in a generic case. To show this, we review Odake and Sasaki's construction for ladder operators \cite{Odake:2006pl,Odake:2006jmp,Odake:2016} in the next subsection.

\subsection{Odake and Sasaki's construction for ladder operators}
It was shown in Refs.\ \cite{Odake:2006pl,Odake:2006jmp,Odake:2016} that one can construct ladder operators for a Hamiltonian $H$ if there exists a function $\eta$, called a sinusoidal coordinate, which satisfies the closure relation
\begin{align}
	[H, [H,\eta] ] & = [H,\eta] R_1(H) + \eta R_0(H) + R_{-1}(H) \,,
\label{closure_relation}
\end{align}
where $R_0(H)$, $R_1(H)$ and $R_{-1}(H)$ are polynomials in $H$. Given a sinusoidal coordinate $\eta$, the ladder operators are formally provided\footnote{We point out here that the derivation of (\ref{ladd_dc_2}) is independent of the fact that $\eta$ is a function, which implies that if the closure relation (\ref{closure_relation}) is satisfied for a linear operator $\eta$, then $\hat{D}_\pm$ provide higher-order ladder operators.} by
\begin{align}
	\hat{D}_\pm & \equiv [H,\eta] - \eta \alpha_\mp(H) + \frac{R_{-1}(H)}{\alpha_\pm(H)} \,,
\label{ladd_dc}
\end{align}
where $\alpha_\pm$ are given by $R_0=-\alpha_-\alpha_+$ and $R_1=\alpha_+ + \alpha_-$. Indeed, it is shown that
\begin{align}
	[H,\hat{D}_\pm] & = \hat{D}_\pm \alpha_\pm(H) \,,
\label{ladd_dc_2}
\end{align}
which leads to the commutation relation $[H,\hat{D}_\pm] \psi_n= \alpha(E_n)\hat{D}_\pm \psi_n$ for an eigenfunction $\psi_n$ with energy $E_n$, i.e., $H\psi_n=E_n\psi_n$. Thus $\hat{D}_\pm$ are ladder operators that change the energy eigenvalues from $E_n$ to $E_n+\alpha_\pm(E_n)$. It is worth commenting that ladder operators constructed from a sinusoidal coordinate have nothing to do with shape invariance (see also Sect.\ \ref{sec:sip}), so that they cannot factorize the Hamiltonian in general. 

It is difficult in general to express the ladder operators (\ref{ladd_dc}) as differential operators because they consist of the fractional terms $R_{-1}(H)/\alpha_\pm(H)$, and $\alpha_\pm(H)$ are not necessarily polynomials in $H$. To obtain expressions as differential operators, we take the same procedure as that from (\ref{Dhat}) to (\ref{ladder_HO_2}), namely, replace the Hamiltonian $H$ by the energy eigenvalues $E_n$. Actually, when the ladder operators $\hat{D}_\pm$ act on an eigenfunction $\psi_n$ with energy $E_n$, i.e., $H\psi_n=E_n\psi_n$, they are realized by
\begin{align}
	\hat{D}_\pm\psi_n & = \left([H,\eta] - \eta \alpha_\mp(E_n) + \frac{R_{-1}(E_n)}{\alpha_\pm(E_n)} \right)\psi_n \,,
\end{align}
hence we obtain
\begin{align}
	D_\pm(E_n) & \equiv [H,\eta] - \eta \alpha_\mp(E_n) + \frac{R_{-1}(E_n)}{\alpha_\pm(E_n)} \,.
\label{Lad_SO}
\end{align}
Using these expressions, we can explicitly check that $D_\pm(E_n)$ satisfy the commutation relations
\begin{align}
	[H,D_\pm(E_n)] & = \alpha_\pm(E_n)D_\pm(E_n) + Q(n) (H - E_n) \,,
\label{com_rel_Odake-Sasaki}
\end{align}
with some operator $Q(n)$. Thus, we find that ladder operators constructed from a sinusoidal coordinate fit into the commutation relation (\ref{str3}) by replacing the Hamiltonian with the energy eigenvalues. Moreover, Eq.\ (\ref{Lad_SO}) allows us to express the ladder operators as differential operators.

\subsubsection{Calogero--Sutherland model}
%
For example, we shall look at the Calogero--Sutherland model \cite{Calogero:1971,Sutherland:1972} for a single particle, which is a particular case of the P\"{o}schl--Teller potential \cite{Poschl:1933}. The Hamiltonian is given by
\begin{align}
	H_\text{CS}(g) &= -\frac{1}{2}\left[\frac{d^2}{dx^2} -\frac{g(g-1)}{\sin^2x}\right] \nonumber\\
		&= -\frac{1}{2}\left(\frac{d}{dx}+\frac{g}{\tan x}\right) \left(\frac{d}{dx}-\frac{g}{\tan x}\right) + \frac{g^2}{2} \,,
\end{align}
where $g>0$ and $0<x<\pi$. The eigenfunctions and eigenvalues are given by
\begin{align}
	\psi_n(g) &= C_n (\sin x)^g P_n^{(g-1/2,g-1/2)}(\cos x) \,, \\
	E_n(g) &= \frac{1}{2}(n+g)^2 \,, \qquad n=0,1,\dots \,, 
\label{energy-eigenvalue-morse}
\end{align}
where $C_n$ are normalization constants and $P_n^{(\alpha,\beta)}(\xi)$ are the Jacobi polynomials in $\xi$. In Ref.\ \cite{Odake:2006jmp}, it was shown that the sinusoidal coordinate is given by $\eta=\cos x$, and the closure relation (\ref{closure_relation}) becomes
\begin{align}
		[H_\text{CS}, [H_\text{CS},\cos x] ] & = [H_\text{CS},\cos x] + \cos x \left(2H_\text{CS}-\frac{1}{4}\right) \,,
\end{align}
where $R_1=1$, $R_0=2H_\text{CS}-1/4$ and $R_{-1}=0$, which leads to $\alpha_\pm(H_\text{CS})=1/2\pm\sqrt{2H_\text{CS}}$. Hence, the ladder operators (\ref{ladd_dc}) are given by
\begin{align}
	\hat{D}_\pm & = \sin x \frac{d}{dx} \pm \cos x \sqrt{2H_\text{CS}} \,,
\end{align}
where we have used the fact that $[H_\text{CS},\cos x]= (\sin x) d/dx + (\cos x)/2$. Replacing the Hamiltonian by the eigenvalues (\ref{energy-eigenvalue-morse}), we obtain the ladder operators
\begin{align}
	D_\pm(E_n) & = \sin x \frac{d}{dx} \pm (n+g) \cos x \,.
\end{align}
Since $D_\pm(E_n)$ satisfy the commutation relation (\ref{com_rel_Odake-Sasaki}) with $\alpha_\pm(E_n)=1/2\pm (n+g)$ and $Q(n)=2\eta =2\cos x$, they change the eigenvalues from $E_n$ to $E_{n \pm 1}$.

\section{Intertwining operators and parameter shifting}\label{sec:sec_3}
We consider Eq.\ (\ref{top_of_2_1}). Since the parameter $\nu$ can be thought of as parametrizing theories, we may consider the intertwining relation (\ref{intertwining_rel_0}) with $H_1=H(\nu_1)$ and $H_2=H(\nu_2)$ for two values $\nu_1$ and $\nu_2$. More generally, it is reasonable to consider a one-parameter family of intertwining operators $D(\nu)$ depending on $\nu$ by setting $\nu_1=\nu$ and $\nu_2=f(\nu)$, where $f(\nu)$ is a function of $\nu$. The function $f(\nu)$ describes how the parameter $\nu$ in eigenfunctions and eigenvalues is shifted. Thus, the intertwining relation (\ref{intertwining_rel_0}) is modified to the relation
\begin{align}
	H(\nu) D(\nu) & = D(\nu) \Big(H(f(\nu))+\varepsilon(f(\nu))\Big) \,, 
\label{basic_eq}
\end{align}
where we have added the term $\varepsilon(f(\nu))$ to the right-hand side in order to describe shifts of energy eigenvalues. Given this relation, it is shown that, for an eigenfunction $\psi_n(f(\nu))$ for $H(f(\nu))$ with energy $E_n(f(\nu))$, we obtain
\begin{align}
	H(\nu) \Big(D(\nu) \psi_n(f(\nu))\Big)
	& = \Big(E_n(f(\nu))+\varepsilon(f(\nu))\Big) D(\nu)\psi_n(f(\nu)) \,, \label{str_for_ladder_4}
\end{align}
which shows that $D(\nu) \psi_n(f(\nu))$ is an eigenfunction for $H(\nu)$ with energy $E_n(f(\nu))+\varepsilon(f(\nu))$  when $\|D(\nu)\psi_n(f(\nu))\|<\infty$ and $D(\nu)\psi_n(f(\nu))\neq0$. Since $D(\nu)$ map eigenfunctions for $H(f(\nu))$ into eigenfunctions for $H(\nu)$, there exists an integer $m$ for any $n$ such that
\begin{align}
	\psi_{n+m}(\nu)& \propto D(\nu) \psi_n(f(\nu)) \,, \label{lad_act_1}\\
	E_{n+m}(\nu)& = E_n(f(\nu))+\varepsilon(f(\nu)) \,. \label{lad_act_1_2}
\end{align}
Thus, we call such operators that connect eigenfunctions between two Hamiltonians spectral intertwining operators.

Like ladder operators, spectral intertwining operators are also useful to construct full or partial energy spectra. For instance, if $m=1$, $D(\nu_0)$ connects $\psi_n(f(\nu_0))$ to $\psi_{n+1}(\nu_0)$ as shown in Fig.\ \ref{fig:label4}. Hence, $\psi_n(\nu_0)$ for $n=1,2,\dots$ are constructed from $\psi_0(\nu_n)$ by
\begin{align}
	\psi_n(\nu_0)& \propto D(\nu_0)D(\nu_1)D(\nu_2)\dots D(\nu_{n-1})\psi_0( \nu_n ) \,, &
	n & =1,2,\dots,
\end{align}
where $\nu_n=f(\nu_{n-1})=f(f(\nu_{n-2})) = f(f(\cdots f(f(\nu_0))\cdots))$.
The energy spectrum is given by
\begin{align}
	E_n(\nu_0) & = E_0(\nu_n) + \sum_{i=1}^n \varepsilon(\nu_i) \,, &
	n & =1,2,\dots,
\end{align}
In this way, the full energy spectrum can be constructed if the eigenfunctions for the ground states $\psi_0(\nu)$ are obtained for all $\nu$.

\subsection{Shape invariance in supersymmetric quantum mechanics}\label{sec:sip}

In supersymmetric quantum mechanics \cite{Witten:1981}, two Hamiltonians are related to each other by the intertwining relations (\ref{intertwining_relation}). Moreover, if the two Hamiltonians are parametrized by a common parameter $\nu$ and have the shape invariance \cite{Gendenshtein:1983}
\begin{align}
	H_-(\nu) & = H_+(f(\nu)) + \varepsilon(f(\nu)) \,, \label{shape_invariance}
\end{align}
such a system becomes exactly solvable (see, e.g., Ref.\ \cite{Cooper:1995} for a review of supersymmetric quantum mechanics).
We show this in terms of spectral intertwining operators.

Substituting (\ref{shape_invariance}) into the intertwining relations (\ref{intertwining_relation}), we obtain
\begin{align}
	H_+(\nu)a^\dagger(\nu) & = a^\dagger(\nu)\Big(H_+(f(\nu)) + \varepsilon(f(\nu))\Big) \,, 
\label{SUSY_QM_21} \\
	H_+(\nu)a(f^{-1}(\nu)) & = a(f^{-1}(\nu))\Big(H_+(f^{-1}(\nu))-\varepsilon(\nu)\Big) \,,
\label{SUSY_QM_22}
\end{align}
which fit into the spectral intertwining relation (\ref{basic_eq}).
%
\bibliographystyle{apsrev4-1}
\bibliography{note}
%
\newpage%
\end{document}